\documentstyle[12pt]{article}

\title{Quantum Kicked Dynamics and Classical Diffusion}

\author{Marco Frasca \\
        Via E. Gattamelata, 3 \\
        00176 Roma (Italia) }

\date{}

\begin{document}

\maketitle

\newpage

\begin{abstract}

We consider the quantum counterpart of the kicked harmonic oscillator
showing that it undergoes the effect of delocalization in momentum
when the classical diffusional threshold is obeyed. For this case
the ratio between the oscillator frequency and the frequency of the kick
is a rational number, strictly in analogy with the classical
case that does not obey the Kolmogorov-Arnold-Moser theorem as
the unperturbed motion is degenerate.
A tight-binding formulation is derived showing that there is not
delocalization in momentum for irrational ratio of the above frequencies.
In this way, it is straightforwardly seen
that the behavior of the quantum kicked rotator is strictly
similar to the one of the quantum kicked
harmonic oscillator, although the former, in the classical
limit, obeys the Kolmogorov-Arnold-Moser theorem.

\end{abstract}

\newpage

One of the main difficulties facing a theory that tries to understand
classical chaos by quantum mechanics is that no
diffusional behavior of the quantum model appears. This question was
clearly
pointed out in [1-2] by using the quantum version of the kicked rotator
that, classically, gives rise to the standard map.
The latter is the simplest hamiltonian model displaying chaos over a
certain
threshold.
The standard map was generalized in Ref.\ [3] by considering the effect of
a
harmonic oscillator potential,
originating from a constant magnetic field for a charged
particle in a wave packet. The quantized version of this model was
studied quite in depth in Ref.[4], where it was clearly showed, by a
numerical analysis that no localization appears for some
rational value of the kick and the harmonic oscillator frequencies.
A generalization of the above results through a dynamic localization model
was not however derived.

At the classical level, the kicked harmonic
oscillator features two frequencies, the frequency of the oscillator
and that of the kick, while the kicked rotator has just the
frequency of the kick, but while the latter obeys the
Kolmogorov-Arnold-Moser theorem, the former does not since the unperturbed
motion is degenerate [3].
Different behaviors appear at a classical level even if both
models are known to display chaos. The situation
changes at the quantum level as the kicked rotator has the angular
momentum quantized as an integer multiple of the Planck's constant. This
means that the kicked rotator too has now two frequencies, one
being of quantal origin. The main aim of this paper is to show that,
at the quantum level, both models display similar behaviors, while are
known
to differ classically.

The lack of diffusional behavior is what one
should expect when a more fundamental theory tries to
explain the effects of a less fundamental one [5]. This point was
already stressed in Ref.\ [6]. Here, we derive the diffusional limit
of the classical kicked harmonic oscillator by its quantum version and
give for it a tight-binding formulation that can show localization
for an irrational ratio between the frequencies of the model, exactly
as in the case of the quantum kicked rotator.
Then, we can conclude, by strict analogy, that the
quantum analogue of the diffusional behavior of the
classical kicked rotator is the
one with a rational ratio of the two frequencies as for the quantum
kicked harmonic oscillator.

Let us consider a particle in a weak magnetic field and a wave packet. It
was
seen in Ref.\ [3] that the classical hamiltonian can be cast in the form
\begin{equation}
    H = \frac{1}{2}(p^2 + \alpha^2 q^2) -
        \alpha K \sin(q)\sum_{n=-\infty}^{+\infty}\delta(t-nT)
\end{equation}
where the mass of the particle and the wave-number of the kicked
potential is set equal to unity.
Then, $\alpha$ is the frequency of the oscillator, $T$
the period of the kick and $K$ the strength of the perturbation.
We can easily quantize the above hamiltonian by
introducing the ladder operators $a$, $a^+$ obtaining
\begin{equation}
    H = \left(a^+a + \frac{1}{2}\right)\hbar\alpha -
        \alpha K \sin(\beta(a^++a))\sum_{n=-\infty}^{+\infty}\delta(t-nT)
\end{equation}
being now $\beta = \left(\frac{\hbar}{2\alpha}\right)^\frac{1}{2}$. For
one kick the evolution operator is given by [1-2]
\begin{equation}
    U(T^+,0^+) = e^{\left[\frac{i}{\hbar}\alpha K
\sin(\beta(a^++a))\right]}
                 e^{\left[-i\alpha T
\left(a^+a+\frac{1}{2}\right)\right]}.
                 \label{eq:U}
\end{equation}
Setting $|\psi(0^+)> = \sum_n \psi_n(0^+) |n>$ with $|n>$ the eigenstates
of the harmonic oscillator, one gets
$|\psi(T^+)> = e^{\left[\frac{i}{\hbar}\alpha K \sin(\beta(a^++a))\right]}
\sum_n \psi_n(0^+) e^{-i\left(n+\frac{1}{2}\right)\alpha T}|n>$.

It easy to see that we get two different behaviors of the wave function
depending on whether or not the value $\frac{\alpha T}{2\pi}$
is a rational number. For the main resonant case,
$\frac{\alpha T}{2\pi} = 1$, and $|\psi(0^+)> = |0>$ the ground state
of the harmonic oscillator, we arrive at the result
\begin{equation}
   |\psi(T^+)> = -\sum_n J_n(z) |in\beta>
\end{equation}
where $z = \frac{\alpha K}{\hbar}$ and $|in\beta>$ a coherent state
with a pure imaginary parameter. It is not difficult to see that this
wave function agrees, for certain values of time,
with the one of Ref.[6] where
a harmonic oscillator in a plane wave was considered. So, in the following
we use the same argument of [6] for the computation of the diffusional
limit.

It is quite easy to see that, if we turn off the perturbation, since
the particle is in its ground state, it has the best possible
localization in space and momentum. Turning on the perturbation,
while the localization in space is retained, we have
\begin{equation}
    <p^2> = p_0^2(1+2\beta^2 z^2) +
            p_0^2\sum_{\stackrel{m,n}{m\neq n}}
                               [1+(m+n)^2\beta^2]
                               J_m(z)J_n(z)
                               e^{-(m-n)^2\frac{\beta^2}{2}}
\end{equation}
being $p_0 = \left(\frac{\hbar\alpha}{2}\right)^\frac{1}{2}$.
With $\beta \gg 1$, at $\sqrt{2}\beta z \sim 1$ the localization in
momentum is lost, deviating from the original gaussian result. Then,
we easily derive $2\beta^2 z \sim \sqrt{2}\beta \gg 1$ which is
the classical diffusional limit, $K \gg 1$, as in Ref.\ [3].
However, after $N$ kicks, we get
$|\psi(NT^+)> = -\sum_n J_n(Nz) |in\beta>$, so, classical diffusion should
appear also at small $K$ for large $N$.
This, indeed, happens through a stochastic web [3]. It is also easy to
get the result that $<p^2> \propto N^2$ for $N\rightarrow\infty$ as
for the quantum kicked rotator.

Now, we show that the above results could be kept for any rational
value of the ratio between the oscillator frequency and the kicking
frequency.
In fact the model can be put in the form of a tight-binding model typical
of an electron on a one-dimensional lattice, as also happens for the
quantum kicked rotator. By looking at the Floquet eigenstates of
the operator $U$ in eq.(\ref{eq:U}), $U|\psi_\lambda> = e^{-i\lambda}
|\psi_\lambda>$, if one sets $|\bar{\psi}_\lambda> =
\frac{1}{2}\left[1+e^{\frac{i}{\hbar}\alpha K\sin(\beta (a^++a))}
           \right]|\psi_\lambda>$, for the probability amplitudes
defined through $|\bar{\psi}_\lambda> = \sum_{n=0}^{\infty} c_n|n>$, with
$H_0|n>=\left(n+\frac{1}{2}\right)\hbar\alpha|n>$, we arrive at the
tight-binding model
\begin{equation}
    T_n c_n + \sum_{m\neq n} W_{nm}c_m = \epsilon c_n
\end{equation}
being
\begin{equation}
    T_n = \tan\left[\frac{1}{2}\left(\left(n+\frac{1}{2}\right)\alpha T
                   -\lambda\right)\right] \label{eq:tan}
\end{equation}
and
\begin{equation}
    W_{nm} = <n|\tan\left[\frac{\alpha K}{2\hbar}\sin(\beta
(a^++a))\right]|m>
\end{equation}
then, $\epsilon = -W_{nn}$. For this kind of model we have the
standard reference [7] where it is shown that, for
$\frac{\alpha T}{2\pi}$ an irrational number, eq.(\ref{eq:tan}) is
a pseudo-random number generator and we
have Anderson localization. For a rational ratio we have delocalized
Bloch waves. So, we are arrived at an identical situation as for
the quantum kicked rotator. We can conclude that both the quantum
kicked harmonic oscillator and the quantum kicked rotator display
similar behaviors although the latter, in the classical limit, obeys
the Kolmogorov-Arnold-Moser theorem and the former does not. We then
also conclude that a correct description of the quantum analog of
the classical diffusion in the quantum kicked rotator is given by the
case of the rational ratio between the frequencies
$\frac{\hbar}{2I}$ of the free motion (I is the moment of inertia)
and $\frac{2\pi}{T}$ of the kicks.

Several interesting questions are opened up by the above discussion. We
reversed the role of the standard map [1-2] and the standard map with
a twist [3] in the quantum limit. However, our analysis requires more
study on the quantum kicked harmonic oscillator. In fact, a general
Floquet map should be derived in the rational case, as was already done
for
the quantum kicked rotator [8].
A quantum master equation should
be derived for the above cases and the limit $\hbar\rightarrow 0$ taken,
proving that the Fokker-Planck equation of the classical case is obtained.
We conclude by saying that a lot of new interesting physics could arise
from these studies of quantum mechanics, even though quantum theory is now
a well known subject, apart from interpretation matters.

\newpage

[1] G. Casati, B.V.Chirikov, F.M.Izrailev, J.Ford in Lectures Note in
Physics, {\bf 93}, 334 (1979) (Springer-Verlag, Berlin, 1979)

[2] L.E. Reichl,
{\sl The Transition to Chaos (In Conservative Classical Systems:
Quantum Manifestations)} (Springer-Verlag, New York, 1992)

[3] A.A. Chernikov, R.Z. Sagdeev, D.A. Usikov, G.M. Zaslavsky,
Computers Math. Applic. {\bf 17}, 17 (1989); V.V. Afanasiev,
A.A. Chernikov, R.Z. Sagdeev, G.M. Zaslavsky, Phys. Lett. A
{\bf 144}, 229 (1990); see also Ref.[2] and
R.Z. Sagdeev, D.A. Usikov, G.M. Zaslavsky, {\sl Nonlinear Physics}
(Harwood Academic Publishers, Philadelphia, 1992)

[4] G.P.Berman, V.Yu.Rubaev, G.M.Zaslavsky, Nonlinearity {\bf 4}, 543
(1991);
see also G.P.Berman, G.M.Zaslavsky in {\sl Quantum Chaos} edited by
G.Casati and B.Chirikov, (Cambridge University Press, Cambridge, 1995)

[5] P.A.M. Dirac, {\sl The Principles of Quantum Mechanics} (Oxford
University Press, London, 1957)

[6] M. Frasca, Phys. Rev. E {\bf 53}, 1236 (1996)

[7] D.R. Grempel, R.E. Prange, S. Fishman, Phys. Rev. A {\bf 29}, 1639
(1984)

[8] F.M. Izrailev, D.L. Shepelyanski, Theor. Math. Phys. {\bf 43}, 553,
(1980)

\end{document}